\documentclass[a4paper]{jpconf}
\usepackage{graphicx}
\usepackage{amssymb}

\def\slashchar#1{\setbox0=\hbox{$#1$}
   \dimen0=\wd0 \setbox1=\hbox{/} \dimen1=\wd1
   \ifdim\dimen0>\dimen1 \rlap{\hbox to \dimen0{\hfil/\hfil}} #1
   \else  \rlap{\hbox to \dimen1{\hfil$#1$\hfil}} / \fi}

\begin{document}
\title{EM vs Weak Structure Functions in DIS processes}
\author{M. Sajjad Athar, H. Haider}
\address{Department of Physics, Aligarh Muslim University, Aligarh, India 202002}
\ead{sajathar@gmail.com}

\author{I Ruiz Simo, M.J. Vicente Vacas}
\address{Departamento de F\'\i sica Te\'orica e IFIC, Centro Mixto
Universidad de Valencia-CSIC, Institutos de Investigaci\'on de
Paterna, Apartado 22085, E-46071 Valencia, Spain}
\begin{abstract}
We obatin the ratio $F_i^A/F_i^{D}$(i=2,3,~ A=Be, C, Fe, Pb;~ D=Deuteron) in the case of weak and electromagnetic nuclear structure functions. 
For this, relativistic nuclear spectral function which incorporate the effects of Fermi motion, binding and nucleon correlations is used. 
We also consider the pion and rho meson cloud contributions and shadowing and antishadowing effects.
\end{abstract}
\section{Introduction}
In this paper, we  study nuclear medium  effects on the electromagnetic $F^{EM}_2$($x$,$Q^2$)~\cite{Sajjad1} and weak $F_2$($x$,$Q^2$) and $F_3$($x$,$Q^2$)~\cite{prc84, prc85} structure functions, in some nuclear 
targets. We use a relativistic nuclear spectral function~\cite{FernandezdeCordoba:1991wf} to describe the momentum distribution of nucleons in the nucleus and 
define everything within a field theoretical approach where, nucleon propagators are written in terms of this spectral function. The spectral function has been calculated using the 
Lehmann's representation for the relativistic nucleon propagator and nuclear many body theory is used for calculating it for an interacting Fermi sea in nuclear matter. Local density approximation is
then applied to translate these results to finite nuclei~\cite{Sajjad1, Marco}. We have assumed the Callan-Gross relationship for nuclear structure functions ${F_2}^A(x)$ and
 ${F_1}^A(x)$. The contributions of the pion and rho meson clouds  are taken into account in a many body field theoretical approach which is  
based on Refs.~\cite{Marco,GarciaRecio:1994cn}. We have taken into account target mass correction following Ref.~\cite{schienbein} which has significant effect at low $Q^2$, 
moderate and high Bjorken $x$. For the shadowing effect which is important at low $Q^2$ and low x, and modulates the contribution of pion and rho cloud contributions, 
we have followed the works of Kulagin and Petti~\cite{Petti}. We have compared the numerical results 
for the case of electromagnetic structure function $R_{F_2}^{EM,~A}(x,Q^2)=\frac{2F_2^{EM,~A}(x,Q^2)}{AF_2^{EM,~D}(x,Q^2)}$,
with JLab~\cite{Seely} and SLAC~\cite{Gomez:SLAC} data. We have also presented these ratios for weak structure functions i.e. 
$R_{F_i}^{A}(x,Q^2)=\frac{2F_i^{A}(x,Q^2)}{AF_i^{D}(x,Q^2)}$ (i=2,3).

The expression for the electromagnetic nuclear structure function $F_2^{EM,~A}$ for an isoscalar nuclear target is written as~\cite{Sajjad1}:
\begin{eqnarray}\label{F2A_Kulagin}
F_2^{EM,~A}&=&4\int d^3r\int \frac{d^3p}{(2\pi)^3}\frac{M}{E(\mathbf{p})}\int_{-\infty}^\mu d\omega\;S_h(\omega,\mathbf{p},\rho(\mathbf{r}))\frac{\left(1-\gamma\frac{p_z}{M}\right)}{\gamma^2}\nonumber\\ 
&& \times \left(\gamma^2+\frac{6x^2(\mathbf{p}^2-p^2_z)}{Q^2}\right)F_2^{EM,~N}(x_N,Q^2)
\end{eqnarray}
where $F_2^{EM,~N}$ is the free nucleon structure function expressed in terms of nucleon Parton Distribution Functions(PDFs), $S_h$ is the hole spectral function taken from the works of 
Ref.\cite{FernandezdeCordoba:1991wf}, 
\begin{equation}	\label{gamma}
\gamma=\frac{q_z}{q^0}=
\left(1+\frac{4M^2x^2}{Q^2}\right)^{1/2}\,,~~  x_N=\frac{Q^2}{2(p^0q^0-p_zq_z)}.
\end{equation}

The expressions for weak nuclear structure functions $F^A_2(x)$ and $F^A_3(x)$ for nonisoscalar nuclear target are written as~\cite{prc85}:
\begin{eqnarray}\label{f2Anuclei}
F^A_2(x_A,Q^2)&=& 2\int d^3r\int\frac{d^3p}{(2\pi)^3}\frac{M}{E(\mathbf{p})}\left[ \int^{\mu_p}_{-\infty}dp^0\; S^{p}_{h}(p^0,\mathbf{p},k_{F,p}) F_2^{p}(x_N,Q^2) \right. \nonumber   \\ 
 &+& \left. \int^{\mu_n}_{-\infty}dp^0\; S^{n}_{h}(p^0,\mathbf{p},k_{F,n}) F_2^{n}(x_N,Q^2) \right] \frac{x}{x_N} \left(1+\frac{2x_N p_x^2}{M\nu_N}\right),  
\end{eqnarray}
\begin{eqnarray}\label{f3Anuclei}
F_3^A(x_A,Q^2)&=& 2\int d^3r\int\frac{d^3p}{(2\pi)^3}\frac{M}{E(\mathbf{p})}\left[\int^{\mu_p}_{-\infty}dp^0\; S^{p}_{h}(p^0,\mathbf{p},k_{F,p}) F_3^{p}(x_N,Q^2) \right. \nonumber \\
 &+& \left. \int^{\mu_n}_{-\infty}dp^0\; S^{n}_{h}(p^0,\mathbf{p},k_{F,n}) F_3^{n}(x_N,Q^2) \right] \frac{p^0\gamma-p_z}{(p^0-p_z\gamma)\gamma}, 
\end{eqnarray}
where $F_{2,3}^{p}$ and $F_{2,3}^{n}$ are the proton and the neutron structure functions. We have considered separate distributions of Fermi sea for
protons and neutrons. $S^{p}_{h}$ and $S^{n}_{h}$ are the two different spectral functions, each one of them is normalized to the number of protons or
neutrons in the nuclear target. The deuteron structure functions have been calculated using the same formulae as in Eq.\ref{F2A_Kulagin} 
but performing the convolution with the deuteron wave function~\cite{Lacombe:1981eg} squared instead of the spectral function. 
\section{Results and Discussions}
\begin{figure}
\begin{center}
\includegraphics[height=5.1cm, width=5cm]{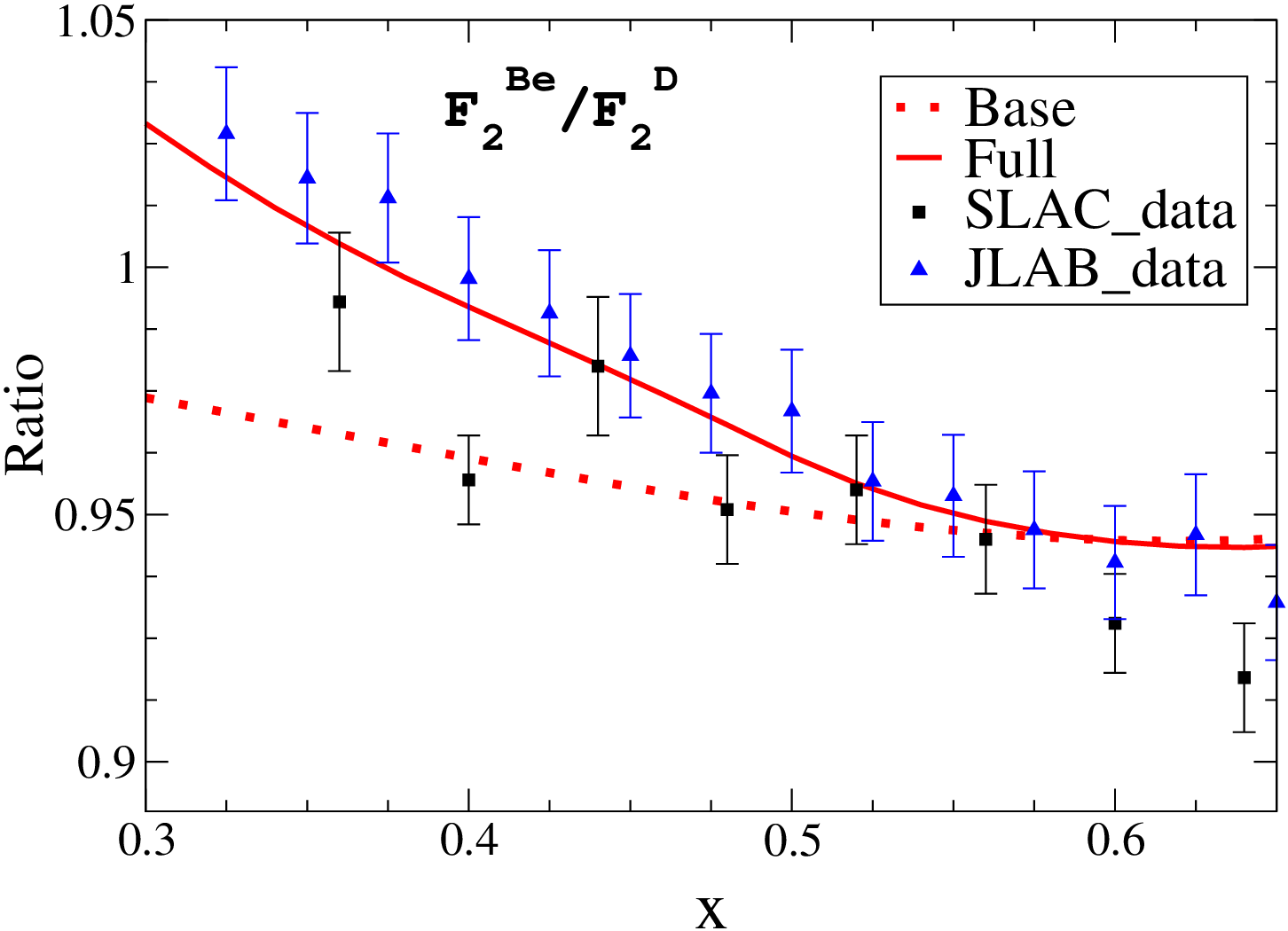}
\includegraphics[height=5.1cm, width=5cm]{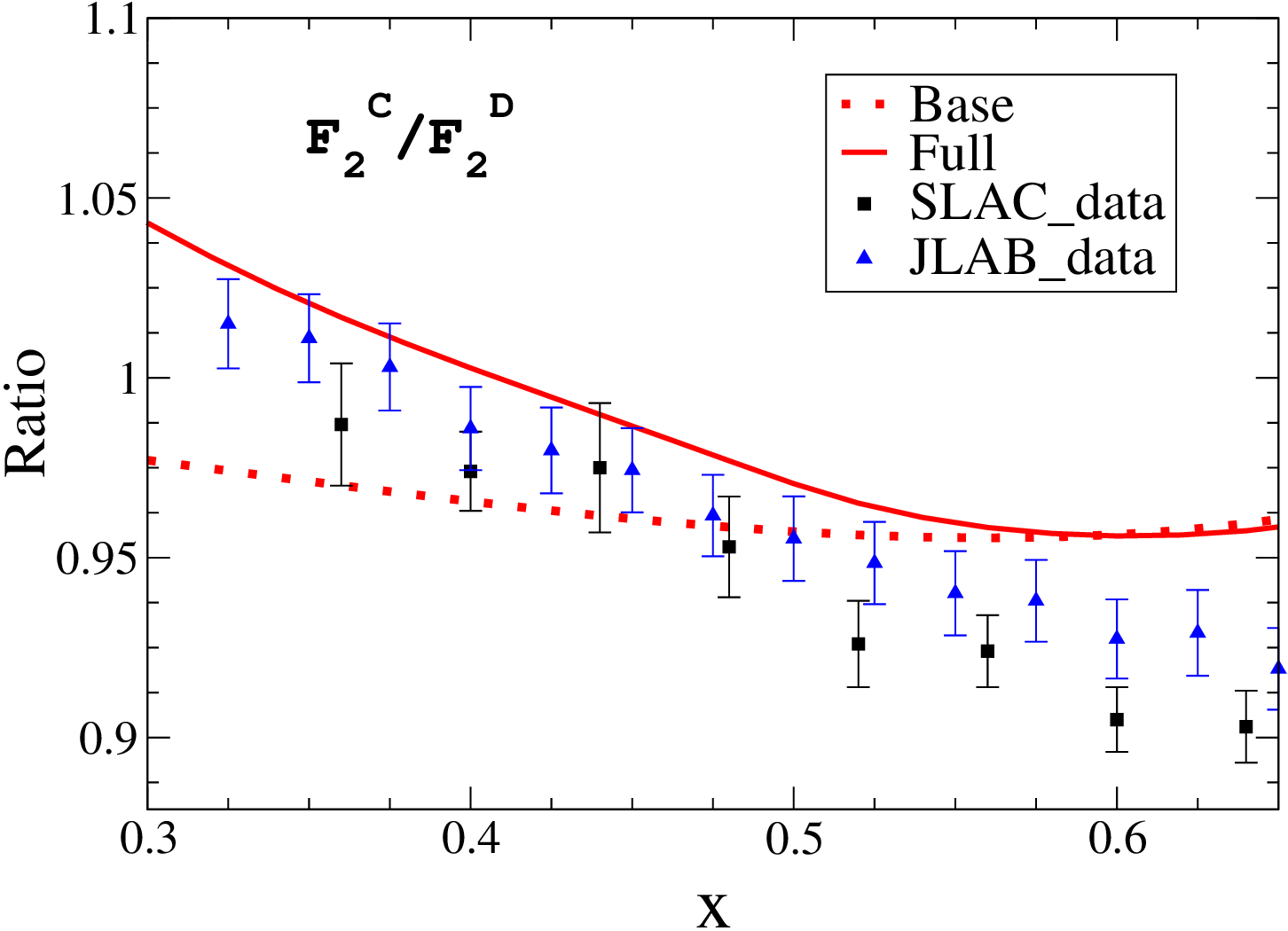}
\includegraphics[height=5.1cm, width=5cm]{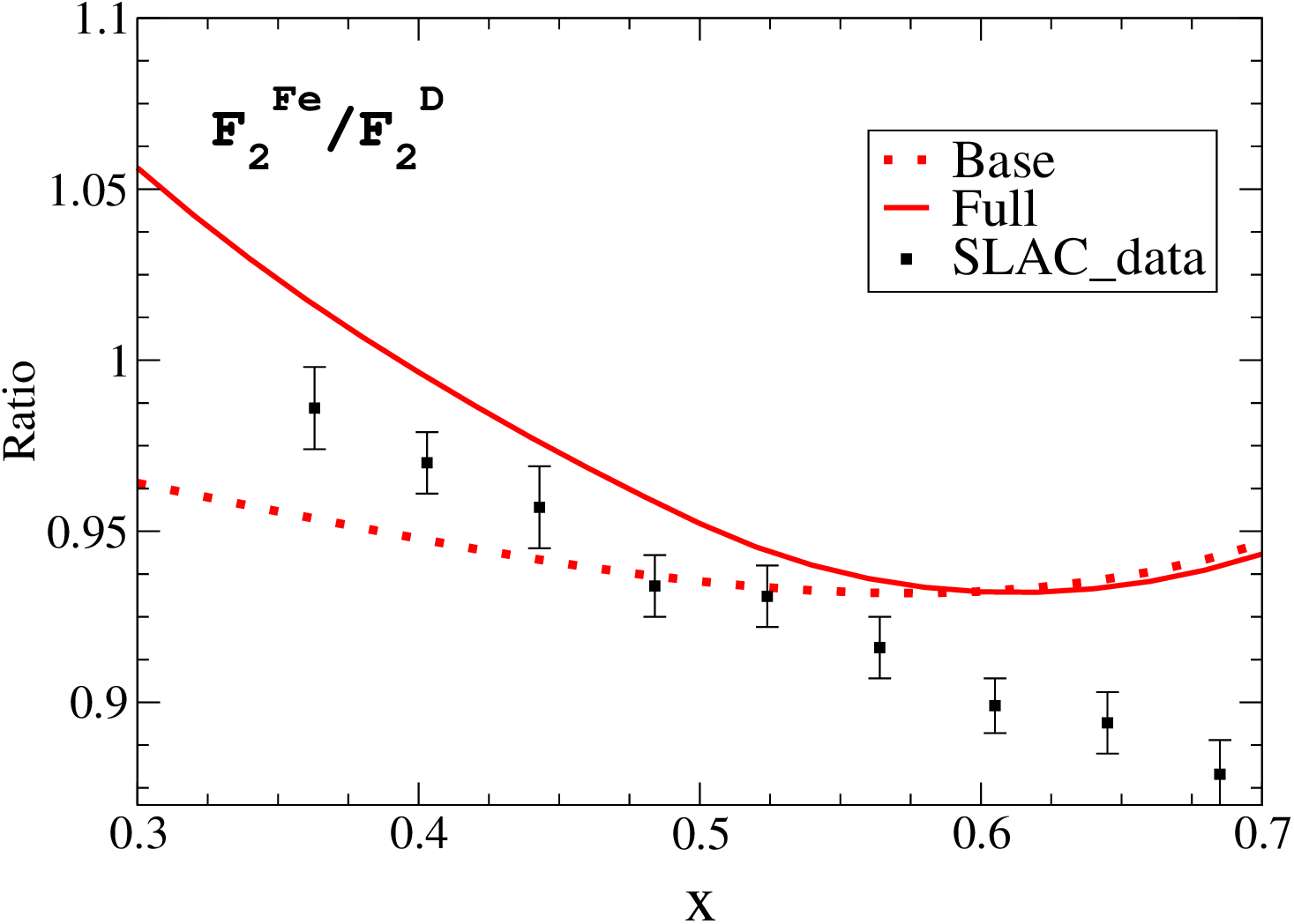}
\caption{$R_{F_2}^{EM,~A}$~vs~$x$~(A=Be, C, Fe). Results with full(base) calculation are shown by solid line(dotted line).}
\label{figEM}
\end{center}
\end{figure}

\begin{figure}
\begin{center}
\includegraphics[height=5.1cm, width=7cm]{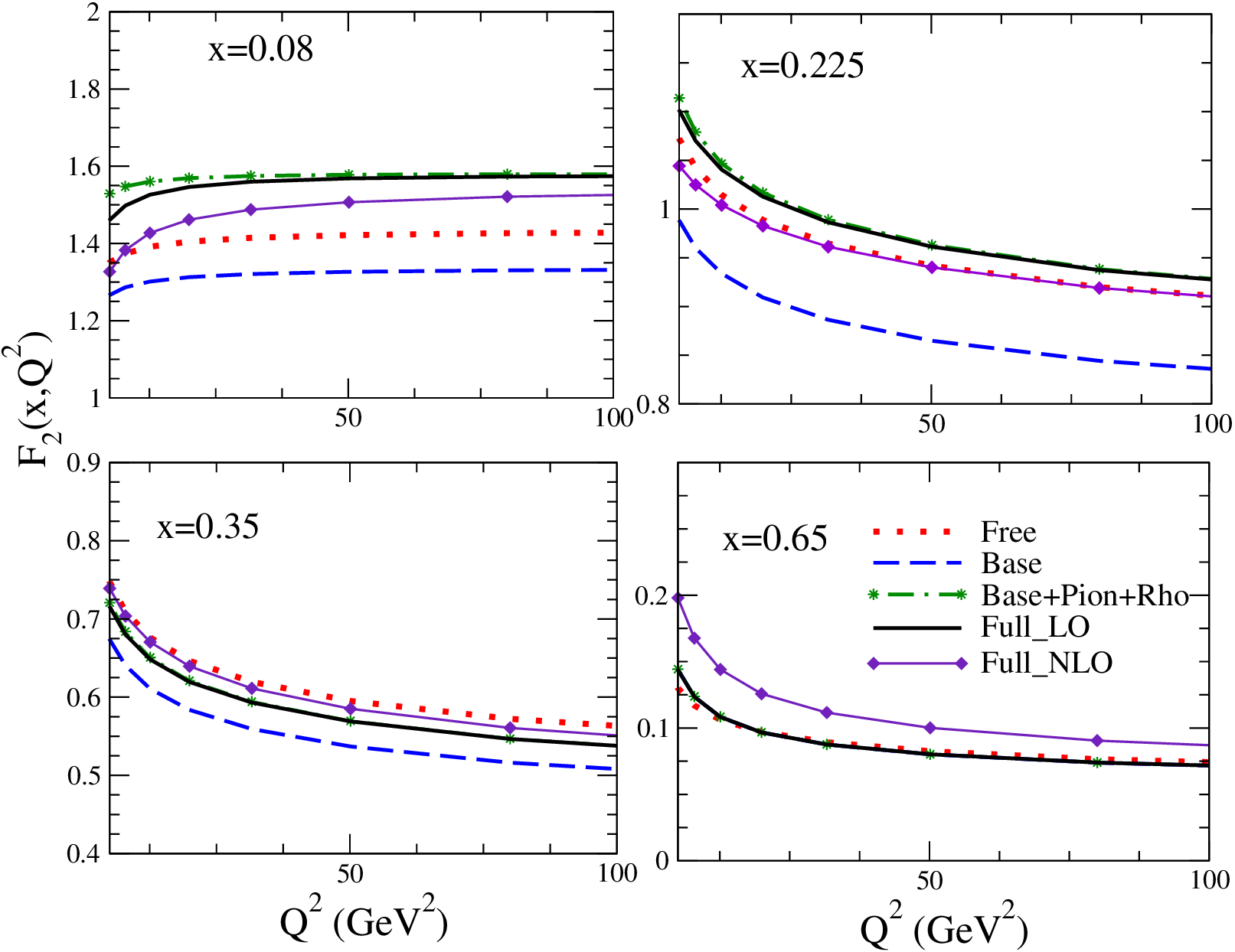}
\includegraphics[height=5.1cm, width=7cm]{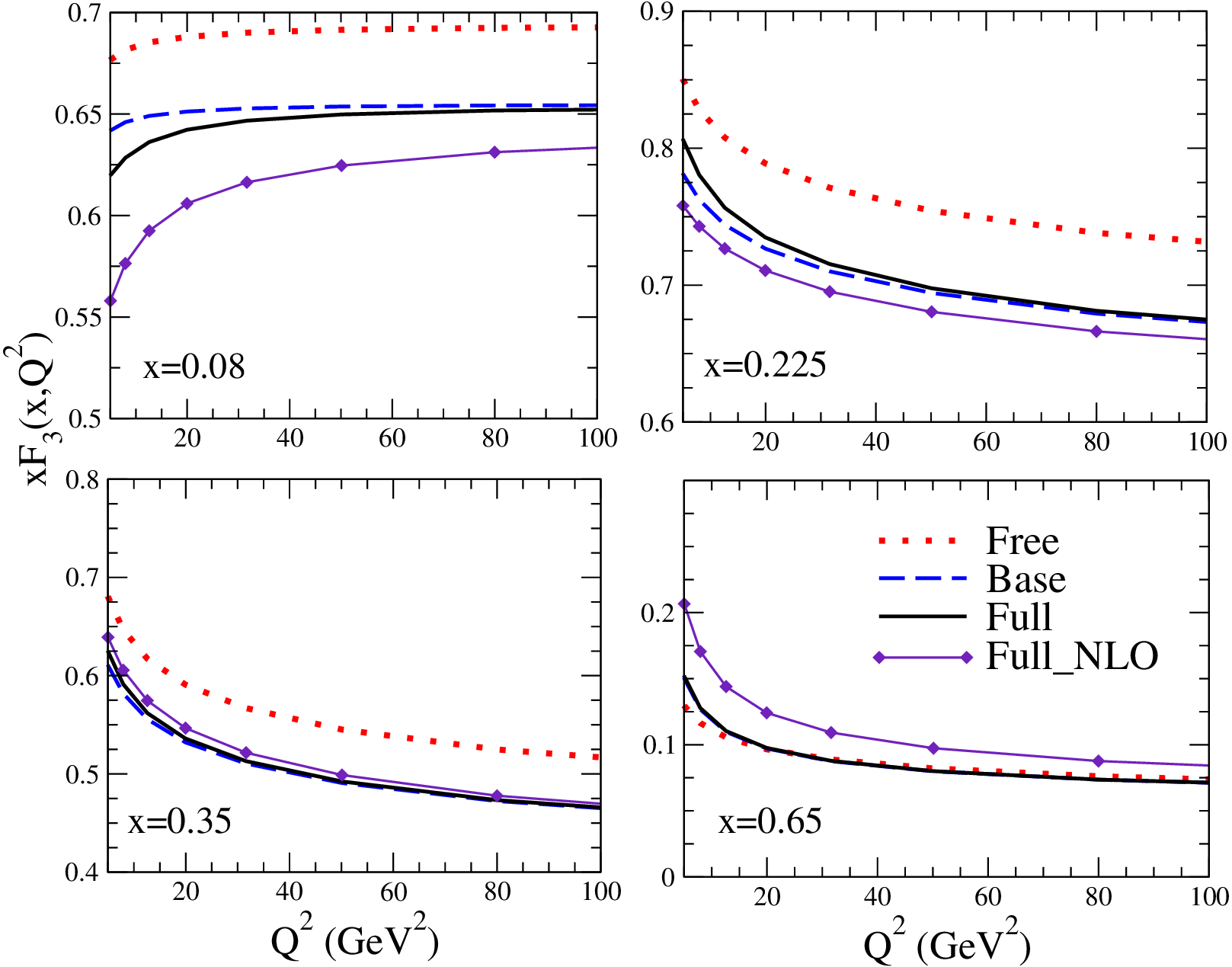}
\caption{Dotted line is $F_i(x,Q^2)$ vs $Q^2$(i=2(Left Panel), i=3(Right Panel)) with no nuclear medium effect. Dashed line is $F_i(x,Q^2)$ vs $Q^2$ in $^{12}C$
obtained by using base results at LO. 
Dotted-dashed line with star is results with no shadowing. Solid line is the full model at LO. Solid line with diamonds is full calculation at NLO.}
\label{f2f3carbon}
\end{center}
\end{figure}

Numerical results obtained by incorporating medium effects like Fermi motion, Pauli blocking, nuclear binding, and nucleon correlations, we call the results as base results.
Results obtained by also including meson cloud contributions and shadowing and antishadowing effects to the base calculation are named as results with the full model calculation.
In Fig.\ref{figEM}, we present the results for the ratio of the electromagnetic nuclear structure funtion to the deuteron structure function. We observe that the
results are in fair agreement with the JLab~\cite{Seely} and SLAC~\cite{Gomez:SLAC} data.
In Fig.\ref{f2f3carbon}, we have presented the numerical results for weak structure functions $F^A_2$ and $F^A_3$ in carbon nuclear target at different values of x and $Q^2$. We find that the base result decreases to 4-6$\%$ from the free case(no medium effects in it) at low-x.
At higher values of x difference between the base and free results vanishes. When the meson cloud contribution is included along with the nucleon spectral function, we see that results change by about 
15-17$\%$ at x=0.08 and modified by 5-6$\%$ at x=0.35, and the difference becomes insignificant at high-x. 
Moreover, we observe that the difference between the results with full model calculation and the results without shadowing and antishadowing effect is about 3-4$\%$ at low-x and low-$Q^2$ which vanishes at high-x.
Also, we find that the results obtained at NLO with full model calculation reduce by about 3-9$\%$ in studied region of $Q^2$ at x=0.08 from
the LO results, and with the increase in x this difference between LO and NLO results, becomes small, like 3-5$\%$ for mid values of x. While at higher values of x the 
results at NLO with full calculation reduce by about 14-24$\%$ from the base results at NLO.
\begin{figure}
\begin{center}
\includegraphics[height=4.8cm, width=5cm]{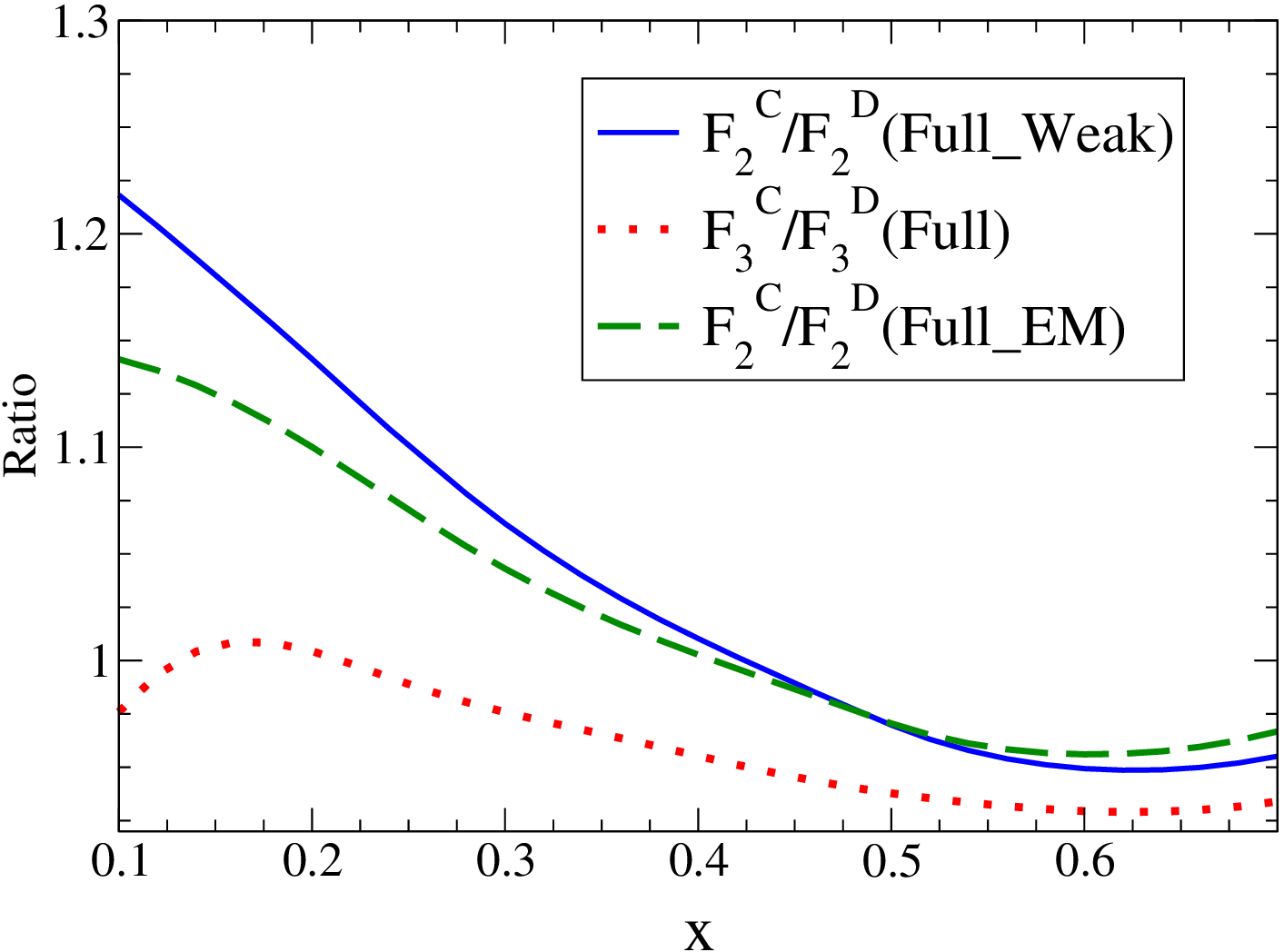}
\includegraphics[height=4.8cm, width=5cm]{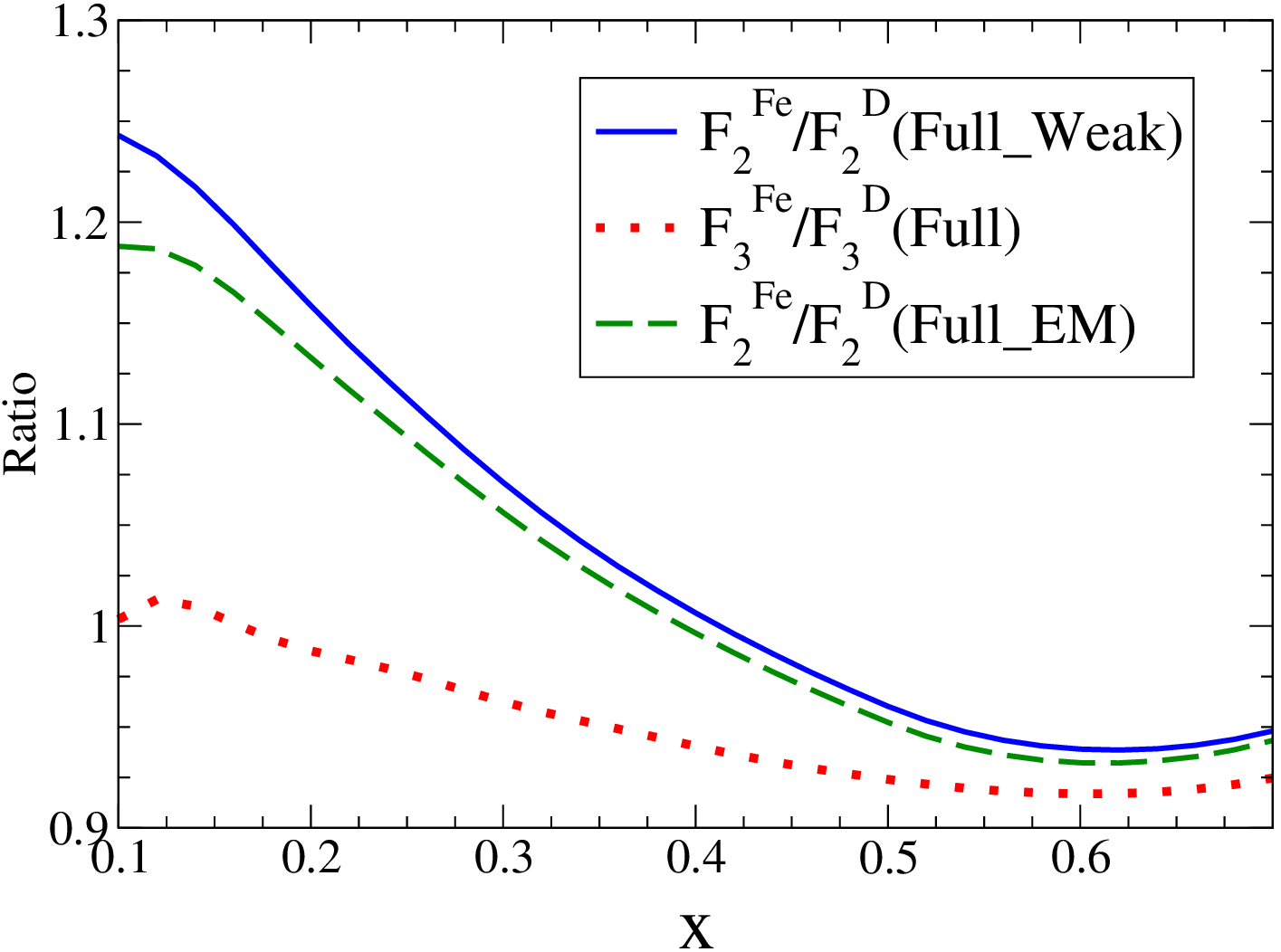}
\includegraphics[height=4.8cm,width=5cm]{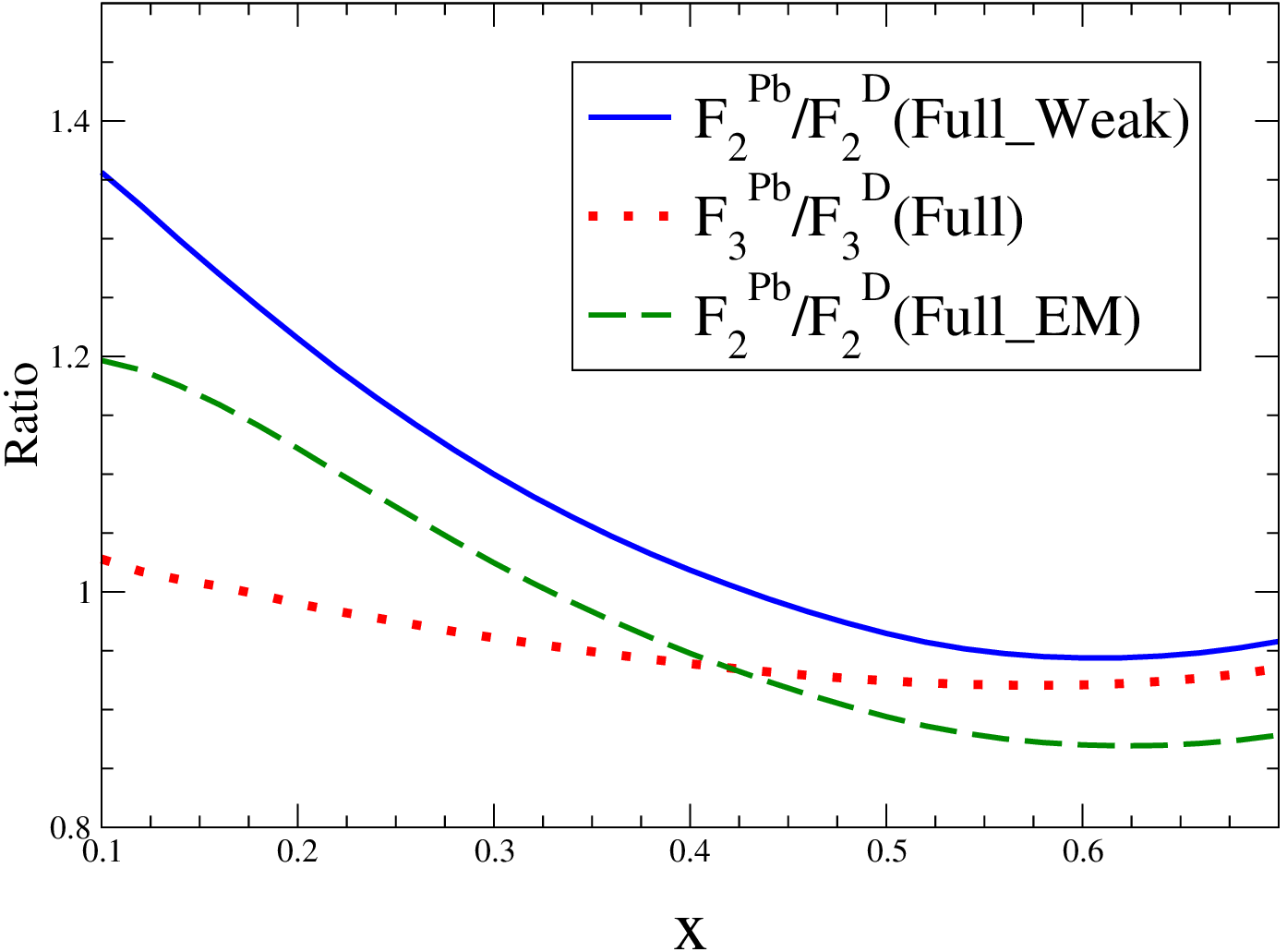}
\caption{Ratio =$\frac{2F_{i}^{A}}{AF_i^D}$(A=C, Fe, Pb and i=2,3).
Dashed line is the result obatined for $R_{F_2}^{EM,~A}$ in the electromagnetic case and solid line is the result
obtained for $R_{F_2}^A$ and dotted line for $R_{F_3}^A$ for the weak case.}
\label{figWeak}
\end{center}
\end{figure}
\begin{figure}
\begin{center}
\includegraphics[height=5.1cm, width=7.8cm]{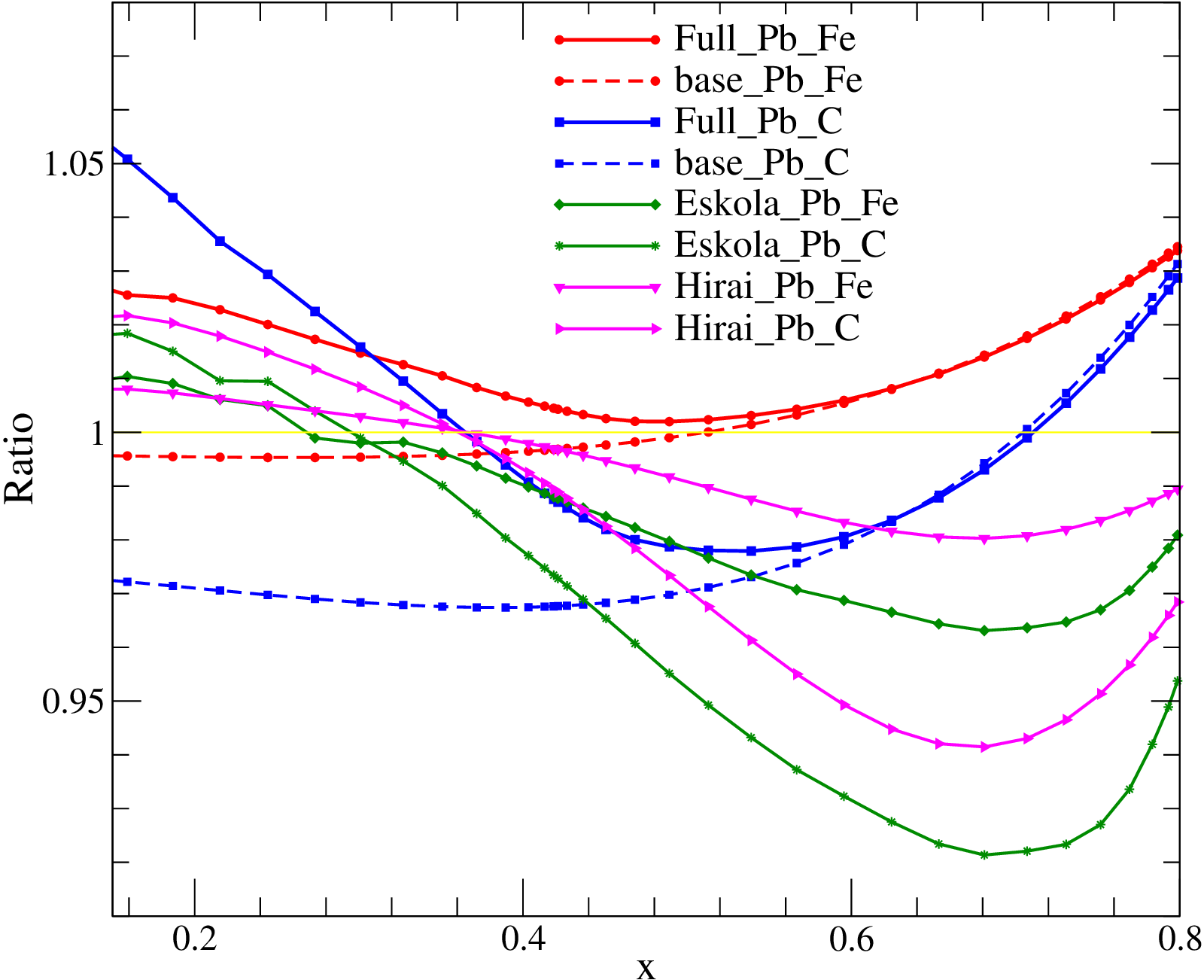}
\includegraphics[height=5.1cm, width=7.8cm]{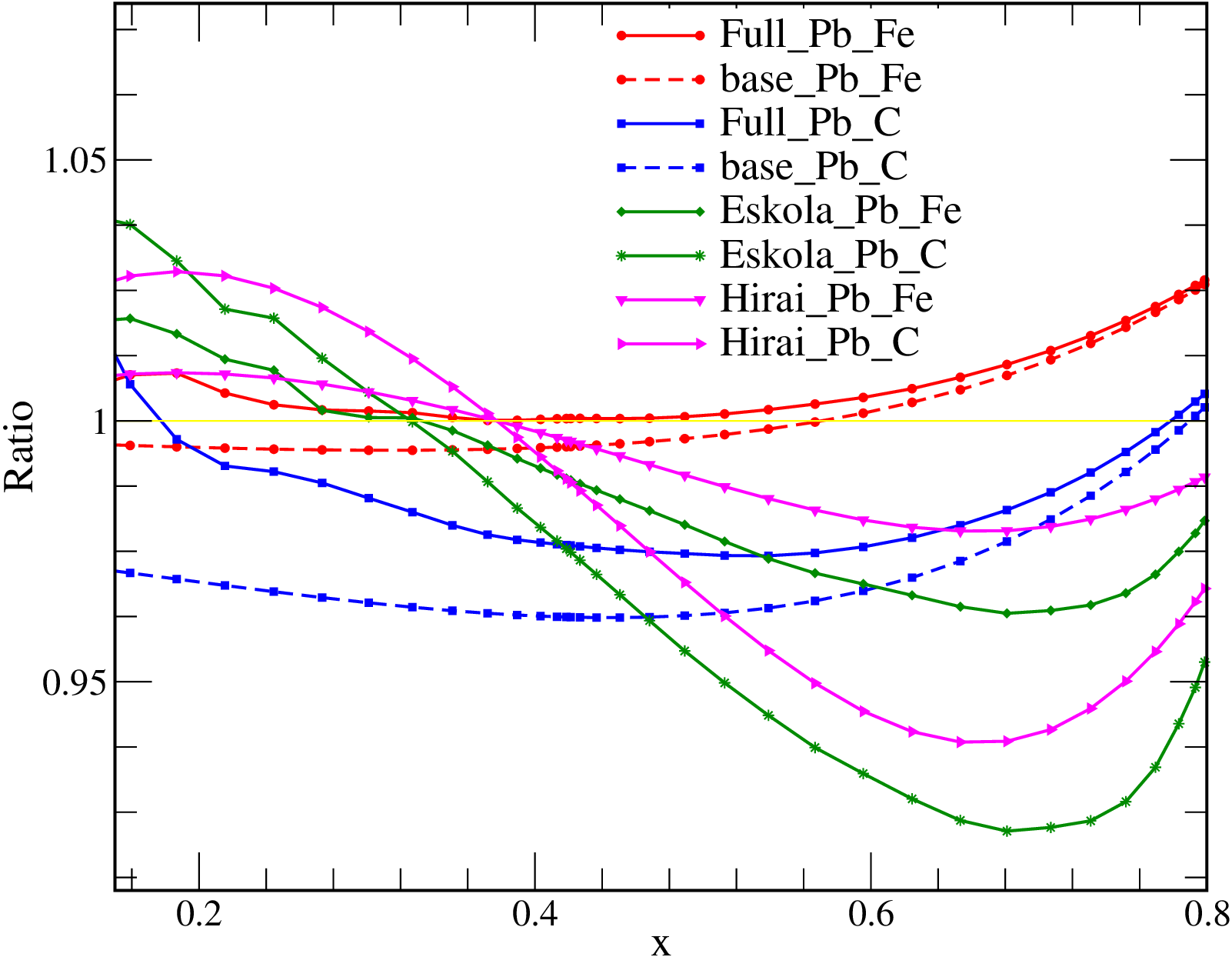}
\caption{Ratio R(x,$Q^2$)=$\frac{12F_{2}^{Pb}}{208F_2^C}$ and R(x,$Q^2$)=$\frac{56F_{2}^{Pb}}{208F_2^{Fe}}$ using our base result(dashed line) and the results obtained
using full model(solid line) at LO for $Q^2=5$ GeV$^2$. The results from 
Hirai et al.~\cite{Hirai} and Eskola et al.~\cite{Eskola} have also been shown.}
\label{figWeakratio}
\end{center}
\end{figure}
In Fig.\ref{figWeak}, the ratio of nuclear structure function obtained in carbon to the deuteron structure function is presented for the weak ($R_{F_i}^{Carbon}, i=2,3$) 
as well as electromagnetic($R_{F_2}^{EM,Carbon}$) case. It may be observed that the nature of the ratio for $F_2$ is different from $F_3$, as well as they are different from the 
electromagnetic case, which is in contrast to many phenomenological studies. MINER$\nu$A~\cite{minerva} is going to study nucleon dynamics in the nuclear medium and their aim is also to 
study weak structure functions. They are taking various nuclear targets like carbon, iron, lead, etc. Therefore, we have studied the ratio of weak structure functions 
 $R_i(x,Q^2)$=$\frac{12F_{i}^{Pb}}{208F_i^C}$ and $\frac{56F_{i}^{Pb}}{208F_i^{Fe}}$ ($i=2,3$) at $Q^2=5$ GeV$^2$ and presented the results in Fig.\ref{figWeakratio}. Here 
 we have also plotted the results obtained using phenomenological prescription of Hirai et al.~\cite{Hirai} and Eskola et al.~\cite{Eskola}. 
 We find the present results to be different from these phenomenological studies. These results may be useful in the analysis of MINER$\nu$A experiment.

One of the authors(MSA) is thankful to PURSE program of D.S.T., Govt. of India and the Aligarh Muslim University for the financial support. This research was supported by the Spanish Ministerio de Economía y Competitividad and
European FEDER funds under Contracts FIS2011-28853-C02-01, by Generalitat Valenciana under Contract No.
PROMETEO/20090090 and by the EU HadronPhysics3 project, Grant Agreement No. 283286.

\end{document}